\documentclass[
reprint,
amsmath,
amssymb, 
aps
]{revtex4-2}%
\usepackage{graphicx}
\usepackage{dcolumn}
\usepackage{bm}
\usepackage{amsmath}
\usepackage{amsfonts}
\usepackage{amssymb}
\usepackage{color}

\begin{document}

\textbf{}
\title{
Effects of successive annealing on high-field electrical transport and the upper critical field in S-substituted FeTe
}

\author{
Ryosuke Kurihara$^1$
}
\email{r.kurihara@rs.tus.ac.jp}
\author{
Satoshi Hakamaka$^1$
}
\author{
Masaki Kondo$^2$
}
\author{
Ryuji Okazaki$^1$
}
\author{
Masashi Tokunaga$^2$
}
\author{
Hiroshi Yaguchi$^1$
}

\affiliation{
$^1$Department of Physics and Astronomy, Faculty of Science and Technology, Tokyo University of Science, Noda, Chiba 278-8510, Japan	
}
\affiliation{
$^2$The Institute for Solid State Physics, The University of Tokyo, Kashiwa, Chiba 277-8581, Japan
}

\begin{abstract}
Since iron-based superconductors have been discovered, many scientists have focused on their characteristic properties, such as an unconventional mechanism and a high upper critical field.
Sulphur-substituted FeTe compounds are one of the members of the iron-based superconductors; however, chemical processes, such as O$_2$ annealing, are needed to induce superconductivity because of the existence of excess iron in as-grown crystals.
Thus, the removal of excess iron and the obtaining of clean sulphur-substituted FeTe can play a key role in the understanding of the superconducting properties and the application to the superconducting devices.
In this study, we present the successive annealing effects on sulphur-substituted FeTe compounds to investigate the electrical transport properties under high magnetic fields.
Our measurements show that successive annealing processes improve the electrical transport properties in the superconducting states under magnetic fields.
The removal of excess iron acting as magnetic impurities is indicated by the improvement of the upper critical field and its analysis.
\end{abstract}

\maketitle

\section{
\label{sect_intro}
Introduction
}
The discovery of high-$T_\mathrm{c}$ superconductors is attracting much attention.
The application of high-temperature superconductors is of interest to many engineers and scientists, and one of the most recent areas of concern may be the restriction on the use of superconducting magnets due to the helium shortage
\cite{Spatolisano_CEM100, Wilkinson_RSEE1}.
Many physicists are probably also fascinated because high-$T_\mathrm{c}$ superconductors are associated with intriguing phenomena, such as quantum criticality
\cite{Lohneysen_RMP79, Stewart_RMP83, Smidman_RMP95},
pseudo-gap
\cite{Lee_RPM78, Imajo_PRM7},
fluctuations due to non-magnetic degrees of freedom
\cite{Fernandes_PRL105, Yoshizawa_JPSJ81, Tazai_JPSJ88, Imajo_PRL125},
etc.
These may be the reasons why much effort has been devoted to searching for new superconductors since the discovery of the first superconductor, Hg, over 100 years ago.
Several strategies are known to discover a high-$T_\mathrm{c}$ superconductor.
One of the ways to discover a new superconductor can be to explore new materials
\cite{Steglich_PRL43, Yagubskii_JETPLett, Bednorz_ZPB64, Kamihara_JACK130}.
Another possibility is to apply high pressure to non-superconducting materials
\cite{Shimizu_JPSJ74}.
Furthermore, the chemical substitution of a non-superconducting material can also be such this way
\cite{Yi_JPSJ82, Tojo_JAP113}.
From the chemical process point of view, annealing can also be an important method for non-superconducting materials.

\begin{figure*}[t]
\begin{center}
\includegraphics[clip, width=1.0\textwidth]{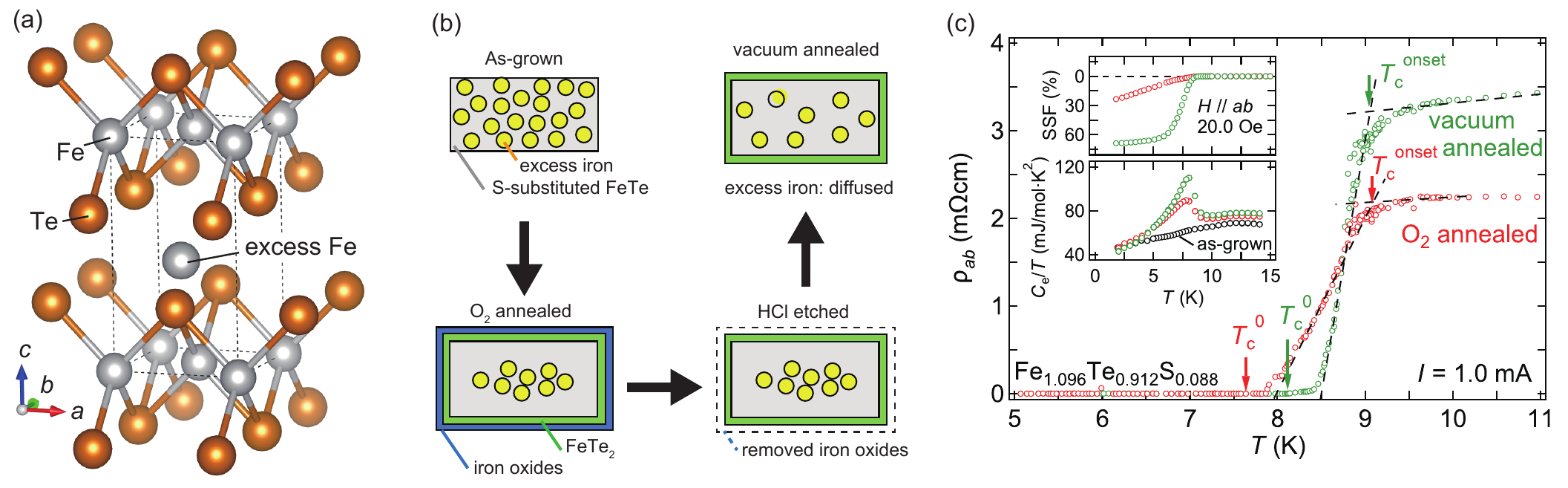}
\end{center}
\caption{
(a) Crystal structure of Fe$_{1+y}$Te with excess iron in the tetragonal phase produced by \textit{VESTA3}
\cite{VESTA}.
The dashed line shows the tetragonal unit cell.
The arrows with labels $a$, $b$, and $c$ indicate the crystallographic orientations.
(b) Schematic diagram of the successive annealing process.
(c) Temperature dependence of the in-plane electrical resistivity $\rho_{ab}$ of the O$_2$- and vacuum-annealed Fe$_{1+y}$Te$_{1-x}$S$_x$ with $x = 0.088$ and $y = 0.096$ measured by a DC current of $I_\mathrm{DC} = 1.0$ mA.
The dashed lines indicate the linear fit of $\rho_{ab}$.
$T_\mathrm{c}^\mathrm{onset}$ and $T_\mathrm{c}^0$ are determined by fitting the slope of $\rho_{ab}$-$T$ and the temperature at which $\rho_{ab}$ shows a zero resistivity value in the experimental resolution, respectively.
The inset in panel (c) indicates the temperature dependence of the superconducting shielding fraction (SSF) of the O$_2$- and vacuum-annealed samples.
The SSF is estimated from the magnetisation under an external magnetic field strength of $ H = 20.0$ Oe applied along the in-plane direction.
The temperature dependence of the electronic specific heat divided by temperature, $C_\mathrm{e} / T$, of the as-grown sample and the O$_2$- and vacuum-annealed samples are also depicted as the inset in panel (c).
Each figure and the data are reproduced from our previous studies
\cite{Kurihara_PRM9}.
}
\label{Fig_Introduction}
\end{figure*}

Sulphur-substituted FeTe compounds, a member of iron-based superconductors, are one of the materials requiring chemical processing.
Figure \ref{Fig_Introduction}(a) shows the crystal structure of the iron chalcogenide Fe$_{1+y}$Te at room temperature.
Here, $y$ denotes the composition of excess iron included in the crystal.
Fe$_{1+y}$Te simultaneously exhibits an antiferromagnetic (AFM) transition and a structural phase transition (SPT) from tetragonal to monoclinic with the space group $P2_1/m$ ($C_{2h}^{2}$, No. 11), but Fe$_{1+y}$Te does not exhibit superconductivity
\cite{Mizuguchi_PhysC469}.
Due to S-substitution in Fe$_{1+y}$Te, denoted as Fe$_{1+y}$Te$ _{1-x}$S$_x$, both the AFM transition and SPT temperatures show lowering.
In Fe$_{1+y}$Te$ _{1-x}$S$_x$, the superconductivity has been observed in polycrystalline samples
\cite{Ma_Vacuum195};
however, no bulk superconductivity has been considered in single crystalline samples in their as-grown state 
\cite{Mizuguchi_JAP109, Mizuguchi_IEEE21}.
To obtain the bulk superconductivity with the superconducting transition temperature of $T_\mathrm{c} \sim 8$ K, processes such as O$_2$ annealing
\cite{Mizuguchi_EPL90},
exposure to air
\cite{Mizuguchi_PRB81},
and soaking in alcoholic beverages
\cite{Deguchi_SST24}
are necessary.
To understand the process-induced superconductivity, the inhomogeneous distribution of the superconducting regions
\cite{Okazaki_JPSJ81},
excess iron deintercalation
\cite{Deguchi_JAP115}
and removal of excess iron due to O$_2$ annealing
\cite{Sun_SciRep4}
have been discussed.

Since these processes realise the bulk superconductivity of Fe$_{1+y}$Te$ _{1-x}$S$_x$, further processes can be required.
The previous studies have proposed that O$_2$ annealing for Fe$_{1+y}$Te$ _{1-x}$S$_x$ induces superconductivity only in the vicinity of the sample surface.
Therefore, we have performed the successive processing to combine O$_2$ annealing, hydrochloric acid (HCl) etching and vacuum annealing [see Fig. \ref{Fig_Introduction}(b)]
\cite{Kurihara_PRM9}
based on the previous studies in Fe$_{1+y}$Te$ _{1-x}$Se$_x$
\cite{Dong_PRM3}.
In this process, excess iron near the sample surface is removed by O$_2$ annealing because of the formation of iron oxide with a small amount of FeTe$_2$.
This iron oxide is removed by HCl etching, and then the residual excess iron is diffused into the whole of the sample due to vacuum annealing.
We have confirmed the emergence of bulk superconductivity due to the removal of excess iron by several physical property measurements, the compositional analysis of excess iron and x-ray diffraction measurements.
We show the temperature dependence of the in-plane electrical resistivity $\rho_{ab}$, the superconducting shielding fraction (SSF) and the electronic specific heat divided by the temperature, $C_\mathrm{e}/T$, as an example of our previous results [see Fig. \ref{Fig_Introduction}(c)].
Compared with $\rho_{ab}$ of the O$_2$-annealed sample, we can see that $\rho_{ab}$ of the vacuum-annealed sample shows a sharp resistive drop and the critical temperature $T_\mathrm{c}^0$, at which $\rho_{ab}$ exhibits zero resistivity, becomes higher.
The emergence of bulk superconductivity has also been indicated by the increase of the SSF and the enhancement of the peak strength of $C_\mathrm{e}/T$.

From the basic research point of view, there can be interest in the investigation of the mechanism of superconductivity of  Fe$_{1+y}$Te$ _{1-x}$S$_x$ using clean samples obtained by the successive annealing.
On the other hand, from the application point of view, it can also be considered important to investigate the potential for application in superconducting magnets.
Therefore, we investigated electrical transport properties under magnetic fields to measure the upper critical field of Fe$_{1+y}$Te$_{1-x}$S$_x$.
Our results can bring about the possibility of application of Fe$_{1+y}$Te$_{1-x}$S$_x$ to superconducting magnets in addition to further understanding of the excess iron contributions to the critical field proposed in the previous study
\cite{Lei_PRB81}.

This paper is organised as follows.
In Sec. \ref{sect_exp}, the sample preparation and experimental procedures are described.
In Sec. \ref{Result and discussion}, we present the temperature dependence of the electrical resistivity under magnetic fields and the high-field magnetoresistivity (MR) of the sample with O$_2$-annealed and vacuum annealed states.
Based on these results, we determined the upper critical field $H_\mathrm{c2}$.
We also discuss the successive annealing effects in the electrical transport properties under magnetic fields.
We conclude our results in Sec. \ref{conclusion}.

\section{
\label{sect_exp}
Experimental details
}

\subsection{
Sample preparations and characterizations
}

Single crystals of Fe$_{1+y}$Te$_{1-x}$S$_x$ with a nominal composition of FeTe$_{0.8}$S$_{0.2}$ were grown by Tamman's method using a homemade furnace
\cite{Kurihara_PRM9}.
The actual compositions of Fe, Te, S, and O of the as-grown crystal were determined by wavelength-dispersive x-ray spectroscopy (WDS) in the electron probe microanalyser with a tungsten filament (JEOL Ltd., JXA-8100) at Research Equipment Center, Tokyo University of Science.
For the quantitative determination of the chemical compositions, FeS$_2$, Te, and SrTiO$_3$ (JEOL Ltd.) were used as standard samples.
An acceleration voltage of 15 kV and a probe current of $2 \times 10^{-8}$ A were used for the quantitative analysis.
We determined the actual atomic ratio Fe$_{1.052 \pm 0.003}$Te$_{0.908 \pm 0.001}$S$_{0.092 \pm 0.001}$ of as-grown samples. 
In the following, $1+y$ and $x$ represent these actual concentrations.

For the successive processing schematically shown in Fig. \ref{Fig_Introduction}(b),
rectangular-shaped crystals with the dimension of $l \times s \times t = 2.01 \times 0.741 \times 0.074$ mm$^3$ were cut with a wire saw.
Here, $l$, $s$, and $t$ denote the length of the long side, short side, and thickness of the sample, respectively.
Atmospheric O$_2$ gas (99.99\%) was then continuously introduced into a furnace (JTEKT THERMO SYSTEMS Co., KTF035N1).
Based on our previous studies, the annealing temperature of 200 $^\circ$C was chosen.
On the other hand, the size of the sample used in the present studies is smaller than that of the previous studies because a resistivity value larger than 0.1 $\Omega$ is required for the high-field measurements in terms of the experimental resolution.
Thus, we chose the O$_2$-annealing time of 20 hours because excessive O$_2$ annealing causes the loss of the superconductivity of S-substituted Fe$_{1+y}$Te
\cite{Yamazaki_JPSJ85}.
200 degrees Celsius-20 hour O$_2$ annealing can also be applicable for the successive processing in terms of superconducting shielding fraction
\cite{Kurihara_PRM9}.
The HCl-etching and vacuum-annealing conditions were the same as in our previous study.
To optimise the annealing conditions, the magnetic moment $m$ along an in-plane magnetic field direction was measured by the Magnetic Property Measurement System (Quantum Design, MPMS) at the Electromagnetic Measurements Section, The Institute for Solid State Physics (ISSP), The University of Tokyo.

\subsection{
Physical property measurements
}

Electrical resistivity was measured by the standard four-contact method.
Ag paste (DuPont, 4922N) and gold wires were used to form electrodes on the sample.
High-field MR measurements were performed at The Institute for Solid State Physics, The University of Tokyo.
The numerical lock-in technique for AC electric signals with a digital storage oscilloscope (Yokogawa, DL850E) and a function generator (Keysight, 33500B), a homemade $^4$He cryostat and a nondestructive pulse magnet with a time duration of 36 ms were used. 
DC and AC electrical resistivity measurements up to 6 T were performed at Research Equipment Center, Tokyo University of Science.
A superconducting magnet, a nanovoltmeter (Keithley, 2182A) for DC measurements and a rock-in-amplifier (Stanford Research Systems, SR830) were used.
The skin depth for AC electric fields with a frequency of 20 kHz, estimated to be 30 mm, was larger than the sample size.
For temperature measurements, a resistance temperature sensor (Lake Shore Cryotronics, Cernox) and temperature controllers (Cryogenic Control Systems, Model 32 or Lake Shore Cryotronics, Model 335) were used.

\subsection{
Analysis
}
To analyse the upper critical fields of Fe$_{1.052 \pm 0.003}$Te$_{0.908 \pm 0.001}$S$_{0.092 \pm 0.001}$, the Julia programming language (ver. 1.9.4) was used.
The source code is available to download from Ref.
\cite{Kurihara_Code}.
For the calculations, SpecialFunctions.jl and Optim.jl packages were used.

\section{
\label{Result and discussion}
Result and discussion
}

\subsection{
\label{Result}
Electrical transport properties
}

\begin{figure*}[thbp]
\begin{center}
\includegraphics[clip, width=1.0\textwidth]{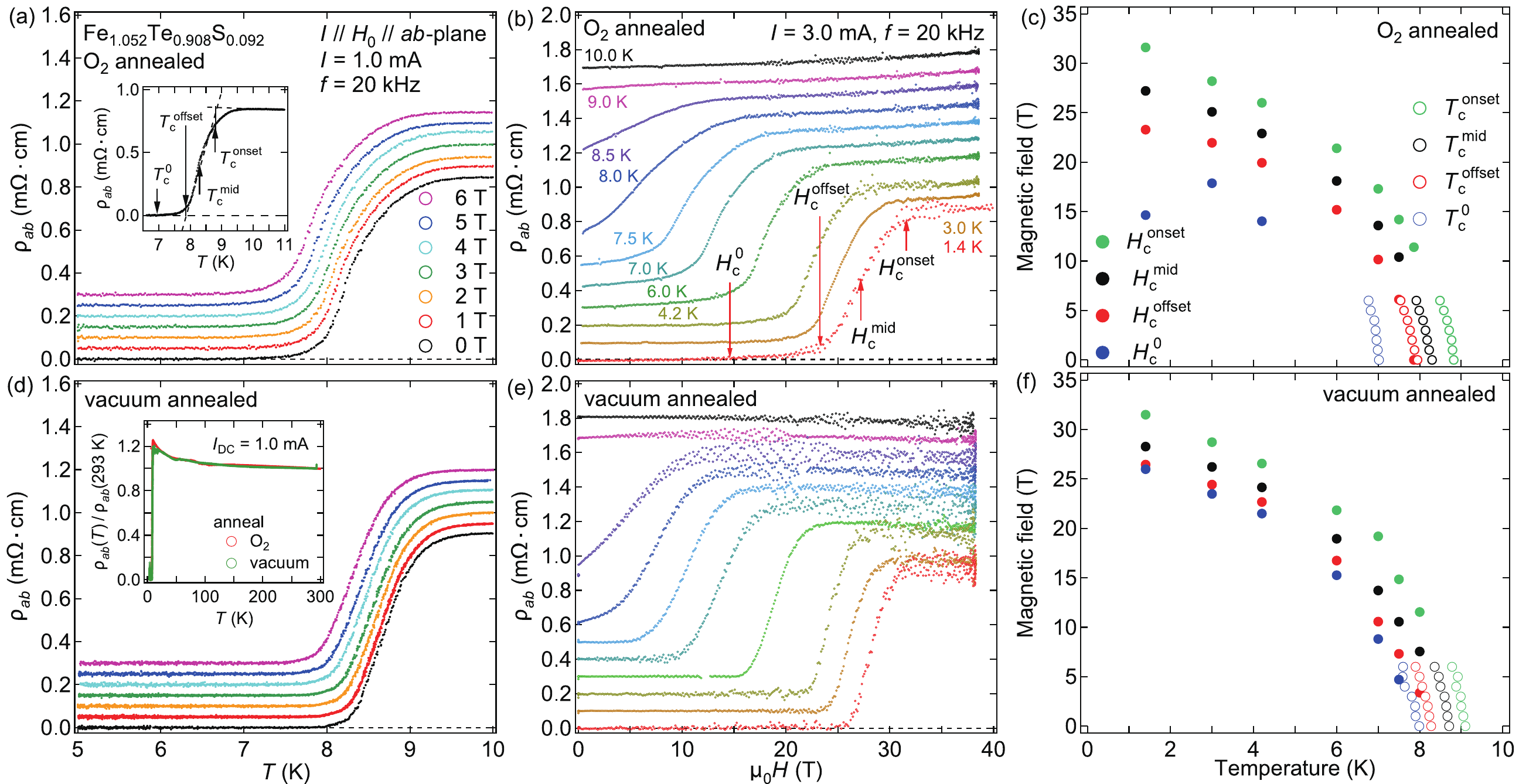}
\end{center}
\caption{
Temperature dependence of the in-plane electrical resistivity $\rho_\mathrm{ab}$ of the (a) O$_2$-annealed [(d) vacuum-annealed] Fe$_{1.052}$Te$_{0.908}$S$_{0.092}$ under several magnetic fields measured by the AC current with the frequency of 20 kHz and the root mean square value of $1.0$ mA.
The external magnetic field $H_0$ is applied along the $ab$-plane of the sample.
$\rho_{ab}$ data under finite magnetic fields in panels (a) and (d) are shifted vertically by 0.1 $\mathrm{m  \Omega \cdot cm}$ for clarity.  
The inset in panel (a) shows a detailed view of the superconducting transition temperatures $T_\mathrm{c}^0$, $T_\mathrm{c}^\mathrm{offset}$, $T_\mathrm{c}^\mathrm{mid}$ and $T_\mathrm{c}^\mathrm{onset}$.
The inset in panel (d) shows the temperature dependence of $\rho_{ab}$ normalized by $\rho_{ab}$ at 293 K of the O$_2$- and vacuum-annealed Fe$_{1.052}$Te$_{0.908}$S$_{0.092}$ at 0 T measured by a DC current of $I_\mathrm{DC} = 1.0$ mA.
Magnetic-field dependence of $\rho_\mathrm{ab}$ of the (b) O$_2$-annealed [(e) vacuum-annealed] Fe$_{1.052}$Te$_{0.908}$S$_{0.092}$ at several temperatures measured by the AC current with the frequency of 20 kHz and the root mean square value of $3.0$ mA.
In each measurement, the magnetic field direction is parallel to the AC current direction of the $ab$-plane.
The red vertical arrows indicate the critical fields $H_\mathrm{c}^0$, $H_\mathrm{c}^\mathrm{offset}$, $H_\mathrm{c}^\mathrm{mid}$ and $H_\mathrm{c}^\mathrm{onset}$.
$\rho_{ab}$ data above 3.0 K in panels (b) and (e) are shifted vertically by 0.1 $\mathrm{m  \Omega \cdot cm}$ for clarity.  
Magnetic field-temperature phase diagram of (c) O$_2$-annealed [(f) vacuum-annealed] Fe$_{1.052}$Te$_{0.908}$S$_{0.092}$.
}
\label{Fig_rhoT}
\end{figure*}

In this section, we present the experimental results of the electrical resistivity measurements of the processed Fe$_{1.052}$Te$_{0.908}$S$_{0.092}$.
We find that vacuum annealing improves the electrical transport properties of the superconducting state under magnetic fields.

To compare the electrical transport properties of the O$_2$- and vacuum-annealing processes, we first measured the O$_2$-annealed Fe$_{1.052}$Te$_{0.908}$S$_{0.092}$.
Figure \ref{Fig_rhoT}(a) shows the temperature dependence of the in-plane electrical resistivity $\rho_\mathrm{ab}$ of the O$_2$-annealed sample under several magnetic fields.
$\rho_{ab}$ shows a resistive drop below the onset of the superconducting transition point of $T_\mathrm{c}^\mathrm{onset} = 8.82$ K.
For further decreasing temperatures, $\rho_{ab}$ becomes zero at $T_\mathrm{c}^0 = 7.02$ K.
As the strength of the applied magnetic field is increased, $T_\mathrm{c}^\mathrm{onset}$ and $T_\mathrm{c}^0$ shift to lower temperatures.
Because the upper critical field $H_\mathrm{c2}$ at low temperatures seemed to be larger than 6 T, we measured the in-plane MR using the pulsed-magnetic field at several temperatures [see Fig. \ref{Fig_rhoT}(b)].
The in-plane MR at 1.4 K shows a finite resistance value above $\mu_0 H_\mathrm{c}^0 = 14.6$ T, then the MR exhibits a rapid increase above $\mu_0 H_\mathrm{c}^\mathrm{offset} = 23.3$ T.
Here, $\mu_0$ is the permeability of vacuum.
Above the onset magnetic field of $\mu_0 H_\mathrm{c}^\mathrm{onset} = 31.6$ T, the MR shows a slight increase in the fields up to 40 T. 
As the temperature is increased, $\mu_0 H_\mathrm{c}^0$ becomes zero around 6 K and $H_\mathrm{c}^\mathrm{offset}$ and $H_\mathrm{c}^\mathrm{onset}$ also show a lowering behaviour.
In addition, the MR near $H_\mathrm{c}^\mathrm{onset}$ becomes broad.

Based on the temperature dependence of $\rho_{ab}$ and the MR, we obtained the magnetic field-temperature phase diagram [see Fig. \ref{Fig_rhoT}(c)].
Here, we plotted $T_\mathrm{c}^\mathrm{mid}$ ($H_\mathrm{c}^\mathrm{mid}$) determined by the half of the electrical resistivity (MR) values at $T_\mathrm{c}^\mathrm{onset}$ ($H_\mathrm{c}^\mathrm{onset}$), namely defined as
$\rho_{ab}\left(T = T_\mathrm{c}^\mathrm{mid} \right) = \rho_{ab}\left(T_\mathrm{c}^\mathrm{onset} \right) / 2$
[
$\rho_{ab}\left(H = H_\mathrm{c}^\mathrm{mid} \right) = \rho_{ab}\left(H_\mathrm{c}^\mathrm{onset} \right) / 2$
] in addition to $T_\mathrm{c}^0$, $T_\mathrm{c}^\mathrm{offset}$ and $T_\mathrm{c}^\mathrm{onset}$ ($H_\mathrm{c}^0$, $H_\mathrm{c}^\mathrm{offset}$ and $H_\mathrm{c}^\mathrm{onset}$).
The upper critical field $H_\mathrm{c2}$ determined by the onset, mid and offset temperatures and magnetic fields shows a monotonic increase with decreasing temperatures.
This behaviour is consistent with the previous studies on sulphur-substituted FeTe compounds
\cite{Lei_PRB81}.
$H_\mathrm{c2}$ determined by $H_\mathrm{c}^\mathrm{onset}$ seems to reach 35 T at $T = 0$ K, however, $H_\mathrm{c2}$ determined by $H_\mathrm{c}^0$ is smaller than 20 T.

\begin{table}[b]
\caption{
The relative change in the critical temperature 
$\mathit{\Delta} T_\mathrm{c}/T_\mathrm{c} = \left( T_\mathrm{c}^\mathrm{onset} - T_\mathrm{c}^0 \right)/ T_\mathrm{c}^\mathrm{onset}$
and the critical field
$\mathit{\Delta} H_\mathrm{c}/H_\mathrm{c} = \left( H_\mathrm{c}^\mathrm{onset} - H_\mathrm{c}^0 \right)/ H_\mathrm{c}^\mathrm{onset}$ 
and their percentage improvement of the O$_2$-annealed and vacuum-annealed Fe$_{1.052}$Te$_{0.908}$S$_{0.092}$.
Abbreviations ann and imp indicate anneal and improvement, respectively.
}
\begin{ruledtabular}
\label{Table_relative}
\begin{tabular}{cccc}

    &$\Delta T_\mathrm{c}/T_\mathrm{c}$ (0 T)
        &$\Delta T_\mathrm{c}/T_\mathrm{c}$ (6 T)
            &$\Delta H_\mathrm{c}/H_\mathrm{c}$ (1.4 K)
\\
\hline
O$_2$ ann
    &20.4\% 
        &20.3\% 
            &53.7\% 
\\
vacuum ann
    &12.2\% 
        &13.4\% 
            &17.5\% 
\\
imp
    &40.2\% 
        &34.0\% 
            &67.4\% 
\end{tabular}
\end{ruledtabular}
\end{table}

After the resistivity measurements on the O$_2$-annealed Fe$_{1.052}$Te$_{0.908}$S$_{0.092}$, we performed the vacuum annealing after the HCl etching on this sample.
Figure \ref{Fig_rhoT}(d) shows the temperature dependence of $\rho_\mathrm{ab}$ of the vacuum-annealed sample under several magnetic fields.
$\rho_{ab}$ shows a resistive drop below $T_\mathrm{c}^\mathrm{onset} = 9.10$ K and a zero resistivity below $T_\mathrm{c}^0 = 7.99$ K.
$T_\mathrm{c}^\mathrm{onset}$ and $T_\mathrm{c}^0$ shift to lower temperatures due to the applied magnetic fields.
Compared with $\rho_{ab}$ of the O$_2$-annealed sample, we confirmed that $\rho_{ab}$ of the vacuum-annealed sample under magnetic fields shows a sharp resistive drop below $T_\mathrm{c}^\mathrm{onset}$ in addition to $\rho_{ab}$ under zero magnetic fields.
As shown in Table \ref{Table_relative}, this improvement can be characterised by the relative variation of the superconducting transition temperature, defined as 
$\mathit{\Delta} T_\mathrm{c}/T_\mathrm{c} = \left( T_\mathrm{c}^\mathrm{onset} - T_\mathrm{c}^0 \right)/ T_\mathrm{c}^\mathrm{onset}$,
which has been considered to describe the quality of superconducting samples
\cite{Wu_PhysC469, Lei_MCP127}.
Similar improvement is also confirmed by the in-plane MR [see Fig. \ref{Fig_rhoT}(e)].
The significant improvements can be that $H_\mathrm{c}^0$ becomes higher.
$H_\mathrm{c}^0$ at 1.4 K changes from $14.6$ T to $26.0$ T due to the successive processing.
As a result of this improvement, $H_\mathrm{c}^0$ is observed even at $7.5$ K, which is just below $T_\mathrm{c}^0 = 7.99$ K.
Although $H_\mathrm{c}^\mathrm{onset}$ does not show a significant improvement, the broadening of the MR around $H_\mathrm{c}^\mathrm{onset}$ is suppressed.
We can confirm this improvement by the relative change in the critical field, defined by
$\mathit{\Delta} H_\mathrm{c} = \left( H_\mathrm{c}^\mathrm{onset} - H_\mathrm{c}^0 \right)/ H_\mathrm{c}^\mathrm{onset}$
[see Table \ref{Table_relative}].
We summarised the characteristic temperatures and fields in the magnetic field-temperature phase diagram [see Fig. \ref{Fig_rhoT}(f)].

As shown above, we confirmed the improvement of the electrical transport properties under high magnetic fields due to successive processing.
Based on the removal of excess iron by the successive processing, our experimental results and the decreasing of the relative change of the critical temperature $\mathit{\Delta} T_\mathrm{c}$ and the critical field $\mathit{\Delta} H_\mathrm{c}$ can reflect the reduction of the impurity contribution to the electrical transport properties under magnetic fields.
If we consider the application of S-substituted Fe$_{1+y}$Te to superconducting magnets, successive processing is probably an effective method.

\subsection{
\label{Discussion}
Analysis and discussion
}

\begin{figure}[t]
\begin{center}
\includegraphics[clip, width=0.45\textwidth]{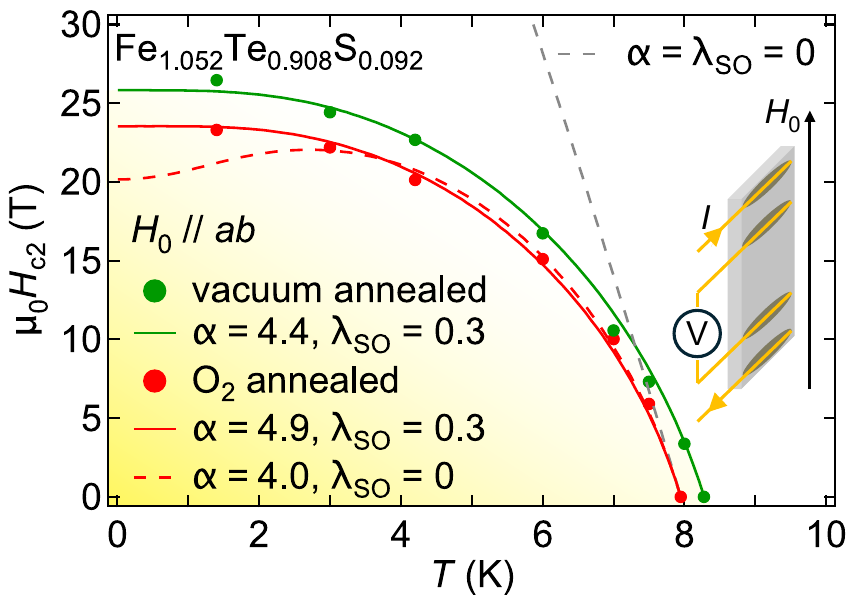}
\end{center}
\caption{
Analytical results of the upper critical field $H_\mathrm{c2}$ of the O$_2$-annealed (red filled circles) and vacuum-annealed (green filled circles) Fe$_{1.052}$Te$_{0.908}$S$_{0.092}$ determined by $H_\mathrm{c}^\mathrm{offset}$ for the in-plane field direction, $H_0 // ab$.
The red (green) solid line indicates the fit of the $H_\mathrm{c2}$ of the O$_2$ (vacuum) annealed sample using Eq. (\ref{single band WHH}) with $\alpha = 4.9$ and $\lambda_\mathrm{SO} = 0.3$ ($\alpha = 4.4$ and $\lambda_\mathrm{SO} = 0.3$).
The red (grey) broken line shows the analytical result with $\alpha = 4.0$ and $\lambda_\mathrm{SO} = 0$ ($\alpha = 0$ and $\lambda_\mathrm{SO} = 0$) for the O$_2$-annealed sample.
The inset shows the schematic view of the electrical resistivity measurements under magnetic fields.
}
\label{Fig_WHH Fit}
\end{figure}

For further understanding of the effect of the successive processing on the electrical transport properties under high-magnetic fields, we focus on the analysis of the upper critical field.
Considering the resistive value, $\rho_{ab} \sim 0.9$ $\mathrm{m \Omega \cdot cm}$ in processed samples just above the superconducting transition temperature and the Hall coefficient $R_\mathrm{H} \sim 5 \times 10^{-9} \mathrm{m^3/C}$
\cite{Ma_Vacuum195},
we can estimated the in-plane mean free path, $l_\mathrm{mean} = \hbar \left( 3 \pi^2 \right)^{1/3} / \left( e^2 \rho_{ab} n^{2/3} \right)$
\cite{Khim_PRB84},
to be $8 \times 10^{-10}$ m.
Here, $\hbar$, $e$ and $n$ are the Dirac constant, elementary charge and the number density, respectively.
Because $l_\mathrm{mean}$ is smaller than the coherence length of superconducting Fe$_{1+y}$Te$ _{1-x}$S$_x$
\cite{Lei_PRB81},
we deduce that the dirty limit description is suitable for the upper critical field.
In addition, the dirty-type superconducting properties have been proposed by the London penetration depth measurements
\cite{Kim_PRB81},
and the characteristic S-shape temperature dependence of the upper critical field in multiband superconductors
\cite{Gurevich_PRB67, Gurevich_PRB82, Gurevich_SuST17}
seems to be absent in our experimental resolutions.
Based on the above, therefore, we focus on the Werthamer-Helfand-Hohenberg (WHH) model
\cite{WHH_PR147}
for dirty limit superconductors.
We stress that similar results have been shown in the previous studies in Fe$_{1+y}$Te$ _{1-x}$S$_x$
\cite{Lei_PRB81}.
We can describe the temperature dependence of the upper critical field by solving the following equation:
\begin{align}
\label{single band WHH}
\ln\frac{1}{t}
= \left(\frac{1}{2} + \frac{i \lambda_\mathrm{SO} }{4 \gamma} \right) \psi \left( \frac{1}{2} + \frac{ \overline{h} +  \frac{1}{2} \lambda_\mathrm{SO}  + i \gamma } { 2 t} \right)
\nonumber   \\
+ \left(\frac{1}{2} - \frac{i \lambda_\mathrm{SO} }{4 \gamma} \right) \psi \left( \frac{1}{2} - \frac{ \overline{h} +  \frac{1}{2} \lambda_\mathrm{SO}  - i \gamma } { 2 t} \right)
- \psi\left( \frac{1}{2} \right)
,
\end{align}
where $t = T/T_\mathrm{c}$,
$\overline{h}
= 4 H_\mathrm{c2} \left( T \right) / \left(- \pi^2 dH_\mathrm{c2} / dt \left|_{t = 1 } \right. \right)$,
$\gamma
= \sqrt{ \left( \alpha \overline{h} \right)^2 - \left( \lambda_\mathrm{SO} /2 \right)^2 }$
and $\psi \left(x \right)$ is the digamma function.
Here, $\alpha$ is the Maki parameter characterising Pauli-spin paramagnetic effect and $\lambda_\mathrm{SO}$ is the spin-orbit coupling contribution
\cite{Maki_PR148}.

\begin{table*}[htbp]
\caption{
Fitting parameters and estimated critical fields obtained by the WHH model of Eq. (\ref{single band WHH}) of the O$_2$-annealed and vacuum-annealed Fe$_{1.052}$Te$_{0.908}$S$_{0.092}$ for the in-plane field direction.
}
\begin{ruledtabular}
\label{Table_parameters}
\begin{tabular}{ccccccccc}

    &$T_\mathrm{c}^\mathrm{WHH}$ (K)
        &$\alpha$
            &$\lambda_\mathrm{SO}$
                &$\mu_0 d H_\mathrm{c2} / dt \left|_{t = 1 } \right.$ (T)
                    &$\mu_0 H _\mathrm{c2}^\mathrm{WHH} \left( 0 \right)$ (T)
                        &$\mu_0 H_\mathrm{c2}^\mathrm{orb} \left(0 \right)$ (T)
                            &$\mu_0 H^\mathrm{P} \left( 0 \right)$ (T)      
                                &$1.84 T_\mathrm{c}$ (T)
\\
\hline
O$_2$ annealed
    &7.95
        &4.9
            &0.3 
                &$-120$
                    &23.5   
                        &83.2   
                            &24.0   
                                &14.6  
\\
vacuum annealed 
    &8.28
        &4.4  
            &0.3
                &$-120$   
                    &25.8   
                        &83.2   
                            &26.7   
                                &15.2  
\\
\end{tabular}
\end{ruledtabular}
\end{table*}

Based on the WHH model, we analysed the temperature dependence of $H_\mathrm{c2}$ of the O$_2$-annealed Fe$_{1.052}$Te$_{0.908}$S$_{0.092}$ for the in-plane magnetic field direction (see Fig. \ref{Fig_WHH Fit}).
For the following discussion, we used $H_\mathrm{c2}$ determined by $H_\mathrm{c}^\mathrm{offset}$.
It should be noted that the offset critical field $H_\mathrm{c}^\mathrm{offset}$ agrees well with $H_\mathrm{c2}$ determined by RF shift measurements
\cite{Mun_PRB83}.
The WHH model with the transition temperature $T_\mathrm{c}^\mathrm{WHH} = T_\mathrm{c
}^\mathrm{offset}$, $\alpha = 4.9$, $\lambda_\mathrm{SO} = 0.3$ and $\mu_0 d H_\mathrm{c2} / dt \left|_{t = 1 } = -120 \right.$ T describes the temperature dependence of $H_\mathrm{c2}$.
We also show that neither the analytical curve with $\alpha = 0$ and $\lambda_\mathrm{SO} = 0$ nor the curve with $\alpha \neq 0$ and $\lambda_\mathrm{SO} = 0$ describe our experimental data.
In addition, we estimated the orbital pair-breaking field at $T = 0$ K, given by
$\mu_0 H_\mathrm{c2}^\mathrm{orb} \left( T = 0 \right) = -0.693 \times \mu_0 d H_\mathrm{c2} / dt \left|_{t = 1 } \right.$, 
the Pauli-limit field at $T = 0$ K, given by 
$\mu_0  H^\mathrm{P} \left( T = 0 \right) = \sqrt{2} \mu_0 H _\mathrm{c2}^\mathrm{orb} \left( T = 0 \right) / \alpha$,
and the Pauli limit field without orbital effect, given by $1.84 T_\mathrm{c}$.
These estimated values are summarised in Table \ref{Table_parameters}.
As shown in the previous study on sulphur-substituted FeTe compounds
\cite{Lei_PRB81},
our results are consistent with the dominant contribution of the paramagnetic effect to the upper critical field of the O$_2$-annealed sample.

In addition to the O$_2$-annealed Fe$_{1.052}$Te$_{0.908}$S$_{0.092}$, we focused on the WHH analysis of the vacuum-annealed sample (see Fig. \ref{Fig_WHH Fit}).
The WHH model describes the temperature dependence of $H_\mathrm{c2}^\mathrm{offset}$.
We stress that the parameters $\lambda_\mathrm{SO}$ and $\mu_0 d H_\mathrm{c2} / dt \left|_{t = 1} \right.$ of the vacuum-annealed sample are the same as those of the O$_2$-annealed sample, while $\alpha$ changes from $4.9$ to $4.4$ (see Table \ref{Table_parameters}).
This result shows that the paramagnetic effect on the upper critical field is suppressed by vacuum annealing while the orbital effect remains unchanged.

To understand the origin of the reduction of the paramagnetic effect due to the successive processing, we focus on the impurity effect. 
As shown in our previous studies
\cite{Kurihara_PRM9},
the removal of excess iron has been revealed by successive processing on sulphur-substituted FeTe compounds.
Based on the previous studies suggesting that the excess iron acts as Kondo-type impurities
\cite{Lei_PRB81},
we can assume that the removal of excess iron reduces the magnetic impurities.
Also, the improvement of the electrical resistivity value due to the successive processing can provide a decrease of $\alpha$ because $\alpha$ is related to the transport lifetime
\cite{Maki_PR148}.
As a result, the paramagnetic effect is suppressed, and then the upper critical field is enhanced from $\mu_0 H_\mathrm{c2}^\mathrm{WHH} \left( 0 \right) = 23.5$ T of the O$_2$-annealed sample to $25.8$ T of the vacuum-annealed sample.
The reduction of impurities can also be consistent with the improvement of the electrical transport properties characterised by the relative variation of the transition temperatures.
Considering the amount of excess iron composition of 0.052 in Fe$_{1.052}$Te$_{0.908}$S$_{0.092}$, approximately one excess iron atom can exist per ten unit cells.
This result is consistent with the ab-plane mean free path, estimated to be $8 \times 10^{-10}$ m, for the lattice parameters of Fe$_{1+y}$Te$ _{1-x}$S$_x$.
The presence of such high-density excess iron can be the origin of large paramagnetic contributions. 
To confirm the above considerations, a comprehensive study of the relationship between $\alpha$ and excess iron composition is necessary. 

In addition to the reduction of the paramagnetic effect, we also discuss the origin of the finite contribution of $\lambda_\mathrm{SO}$ to the upper critical field and its process-independent behaviour.
Based on the theoretical predictions
\cite{Maki_PR148},
the lifetime caused by impurity scattering is related to the impurity density.
In Fe$_{1+y}$Te$ _{1-x}$S$_x$, we deduce that substituted S atoms act as impurities, then a finite value of $\lambda_\mathrm{SO}$ is provided.
Based on our earlier research
\cite{Kurihara_PRM9},
the successive processing does not change the amount of S and Te atoms.
This process independence of S composition is consistent with the conservation of $\lambda_\mathrm{SO}$.
To confirm the above considerations, the S-substitution dependence of $\lambda_\mathrm{SO}$ is required.

Although the improvements by successive processing have been shown for the electrical transport properties, no significant improvement of the upper critical field is observed.
Considering $\rho_{ab}$ normalized by $\rho_{ab}$ at 293 K, denoted as $\rho_{ab}\left( T \right) / \rho_{ab}\left( 293 \mathrm{K} \right) $, of the O$_2$- and vacuum-annealed Fe$_{1.052}$Te$_{0.908}$S$_{0.092}$ under 0 T [see inset in Fig. \ref{Fig_rhoT}(d)], the normal state resistivity exhibits semiconductor-like temperature dependence down to the superconducting transition temperature, indicative of residual excess iron.
On the other hand, suppressed peak values of $\rho_{ab}\left( T \right) / \rho_{ab}\left( 293 \mathrm{K} \right)$ of the vacuum-annealed sample just above the superconducting transition temperature suggest that excess iron is reduced by successive annealing.
Therefore, further reduction of excess iron can enhance the upper critical field of Fe$_{1+y}$Te$_{1-x}$S$_x$.
We believe that further optimisation of the successive processing brings about large upper critical fields comparable to $\mu_0 H_\mathrm{c2}^\mathrm{orb} \left( T = 0 \right)$.

\section{
\label{conclusion}
Conclusion}

We investigated the successive annealing effects in the electrical transport properties of S-substituted Fe$_{1+y}$Te.
To compare the electrical resistivity of the O$_2$-annealed sample with the vacuum-annealed sample, we confirmed that the superconducting state exhibiting a zero resistivity became more robust against magnetic fields due to the successive processing. 
We also confirmed that the upper critical field determined by the MR was improved, but not significantly.
On the other hand, our WHH analysis showed that the paramagnetic contribution to the upper critical field was reduced by successive processing.
Our results indicate that the removal of excess iron due to the successive processing plays a key role in the electrical transport properties and the paramagnetic contributions of S-substituted FeTe compounds.

\section*{Acknowledgment}
We appreciate Keigo Naito, Haruhi Okumura, Yoshifumi Soma, Masaki Daibo and Riara Nakata for their experimental assistance.
This work was partly supported by JSPS transformative research areas (A), section (II) (JP 23H04862, 24H01629).
Our magnetic property measurements were carried out using the facilities of the Materials Design and Characterization Laboratory by the joint research in the Institute for Solid State Physics, The University of Tokyo (No. 202311-MCBXG-0031, 202406-MCBXG-0116, 202411-MCBXG-0042, 202505-MCBXG-0115) and Research Equipment Center, Tokyo University of Science.

\section*{Author contribution}
R.K and S.H. contributed equally to this work.
S.H. and R.K. conceived and designed the experiments.
S.H. performed the sample preparations, annealing and physical property measurements.
R.K. analysed the experimental data.
S.H., R.K. and M.K. performed the high-field MR measurements.
M.T. supervised the conduct of the high-field MR measurements.
S.H. and R.O. measured the electrical resistivity and the MR by the superconducting magnet.
R.K. and S.H. drafted the original manuscript.
H.Y. supervised the conduct of this study.
All authors checked the original manuscript and contributed to its revision.
All authors have agreed to submit the final version of the manuscript.

\bibliography{Reference.bib}

\end{document}